\documentclass[twocolumn]{aastex62}

\usepackage[shortlabels]{enumitem}
\usepackage{upgreek}

\graphicspath{{./}{}}

\received{}
\revised{}
\accepted{} 

\submitjournal{ApJL}

\shorttitle{(6478) Gault: a blue Q-type surface below the dust?}
\shortauthors{Marsset et al.}

\begin{document}

\title{Active asteroid (6478) Gault: a blue Q-type surface below the dust?}

\correspondingauthor{Micha\"el Marsset}
\email{mmarsset@mit.edu}

\author[0000-0001-8617-2425]{Micha\"el Marsset}
\affil{Department of Earth, Atmospheric and Planetary Sciences, MIT, 77 Massachusetts Avenue, Cambridge, MA 02139, USA}

\author[0000-0002-8397-4219]{Francesca  DeMeo}
\affiliation{Department of Earth, Atmospheric and Planetary Sciences, MIT, 77 Massachusetts Avenue, Cambridge, MA 02139, USA}

\author[0000-0001-5465-5449]{Adrian Sonka}
\affiliation{Astronomical Institute, Romanian Academy 5-Cu\c{t}itul de Argint 040557 Bucharest, Romania}
\affiliation{Faculty of Physics, Bucharest University 405, Atomi\c{s}tilor Street 077125 M\v{a}gurele, Ilfov, Romania}

\author[0000-0003-3495-8535]{Mirel Birlan}
\affiliation{IMCCE, Observatoire de Paris, CNRS UMR8028, PSL Research University, 77 avenue Denfert-Rochereau, F-75014 Paris Cedex, France}
\affiliation{Astronomical Institute, Romanian Academy 5-Cu\c{t}itul de Argint 040557 Bucharest, Romania}

\author[0000-0002-6977-3146]{David Polishook}
\affiliation{Department of Particle Physics and Astrophysics, Weizmann Institute of Science, Israel}

\author[0000-0002-6423-0716]{Brian Burt}
\affil{Lowell Observatory, 1400 W. Mars Hill Road, Flagstaff, AZ, 86001, USA}
       
\author[0000-0002-9995-7341]{Richard P. Binzel}
\affiliation{Department of Earth, Atmospheric and Planetary Sciences, MIT, 77 Massachusetts Avenue, Cambridge, MA 02139, USA}

\author[0000-0003-4191-6536]{Shelte J. Bus}
\affiliation{Institute for Astronomy, University of Hawaii, 2860 Woodlawn Drive, Honolulu, HI 96822-1839, USA}

\author[0000-0003-3091-5757]{Cristina Thomas}
\affiliation{Department of Astronomy and Planetary Science, Northern Arizona University, PO Box 6010, Flagstaff, AZ 86005, USA}

\begin{abstract}
We present near-infrared spectroscopy of the sporadically active asteroid (6478)~Gault collected on the 3 m NASA/Infrared Telescope Facility observatory in late 2019 March/early April. 
Long-exposure imaging with the 0.5 m NEEMO~T05 telescope and previously published data simultaneously monitored the asteroid activity, providing context for our measurements.  
We confirm Gault is a silicate-rich (Q- or S-type) object likely linked to the (25)~Phocaea collisional family. 
The asteroid exhibits substantial spectral variability over the 0.75--2.45\,$\upmu$m wavelength range, from unusual blue ($s'$=--13.5$\pm$1.1\%\,$\upmu$m$^{-1}$) to typical red ($s'$=+9.1$\pm$1.2\% $\upmu$m$^{-1}$) spectral slope, that does not   
seem to correlate with activity. 
Spectral comparisons with samples of 
ordinary chondrite meteorites 
suggest that the blue color relates to the partial loss of the asteroid dust regolith, exposing a fresh, dust-free material at its surface. 
The existence of asteroids rotating close to rotational break-up limit and having similar spectral properties as Gault further supports this interpretation. 
Future spectroscopic observations of Gault, when the tails dissipate, will help further testing of our proposed hypothesis. 
\vspace{1cm}

\end{abstract}

\section{Introduction} 
\label{sec:intro}

Cometary-like activity was recently reported on the $\sim$4-km size inner main-belt asteroid (6478)~Gault. 
Tens of millions of kilograms of ejected dust \citep{Hui:2019, Jewitt:2019, Moreno:2019, Ye:2019b} formed a long and thin tail first identified in images collected by the Hawai‘i ATLAS \citep{Tonry:2018} survey in December 2018 \citep{Smith:2019} and subsequently confirmed by other research groups (e.g., \citealt{Hale:2019a,Hale:2019b,Jehin:2019,Lee:2019,Ye:2019a}).
Since then, two additional tails developed \citep{Jewitt:2019}, and archival data back to 2013 revealed earlier phases of activity \citep{Chandler:2019}, ruling out a single collision as the origin of the activity. 
The detection of events near aphelion \citep{Chandler:2019} and the absence of detectable gas in the visible spectrum of Gault \citep{Jewitt:2019} imply volatile sublimation is also unlikely to be responsible. 
On the other hand, signatures of a rotation period of $\sim$2\,hr \citep{Kleyna:2019}, close to the rotational break-up limit \citep{Pravec:2000}, and the very low velocity of the ejected dust grains \citep{Hui:2019, Jewitt:2019, Kleyna:2019, Moreno:2019, Ye:2019b}, indicate that activity may be triggered by rapid spin up due to solar heating (the YORP effect; e.g. \citealt{Bottke:2006,Vokrouhlicky:2015}). 
This remains to be confirmed, as light curves collected so far with various telescopes show very little/no short-term brightness variation \citep{Jewitt:2019,Moreno:2019,Ye:2019b}, possibly because the asteroid is embedded in a dust cloud \citep{Licandro:2000}. 
Likewise, \citet{Ye:2019b} proposed that activity could be triggered by the merging of a near-contact binary, but currently available light curves do not allow this hypothesis to be validated. 

Owing to the lack of spectral measurements before it became active, little is known about the original (pre-outburst) surface composition and taxonomy \citep{Bus:2002, DeMeo:2009} of Gault. 
Dynamically, the asteroid was linked to two overlapping collisional families: Phocaea, which is dominated by S-type asteroids, and Tamara, dominated by low-albedo C-types \citep{Nesvorny:2015, Novakovic:2017}. 
Serendipitous observations from the Zwicky Transient Facility (ZTF) survey \citep{Bellm:2019}, dating back to 2017, returned optical colors similar to C-type or Q-type asteroids \citep{Ye:2019b} (Q-types represent a less-weathered version of the S-type asteroids, e.g. \citealt{Binzel:2010}). 
Additional color measurements collected during active phases between 2018 December and 2019 March returned similar results \citep{Hui:2019, Jewitt:2019, Ye:2019b}. 
In parallel, spectroscopy over the visible and near-infrared (NIR) wavelength range revealed spectral properties that are consistent with Q-type or S-type asteroids (\citealt{Sanchez:2019}, Licandro et al., private communication), supporting Gault's membership to the Phocaea collisional family. 

In this Letter, we report new spectroscopic measurements of Gault measured in the NIR wavelength range between late 2019 March and early April, and near-simultaneous photometry providing context for these measurements. 

\section{Spectroscopy} \label{sec:spectro}

\subsection{Observations and data reduction}

\begin{figure}[ht]
   \centering
   \includegraphics[width=8.5cm]{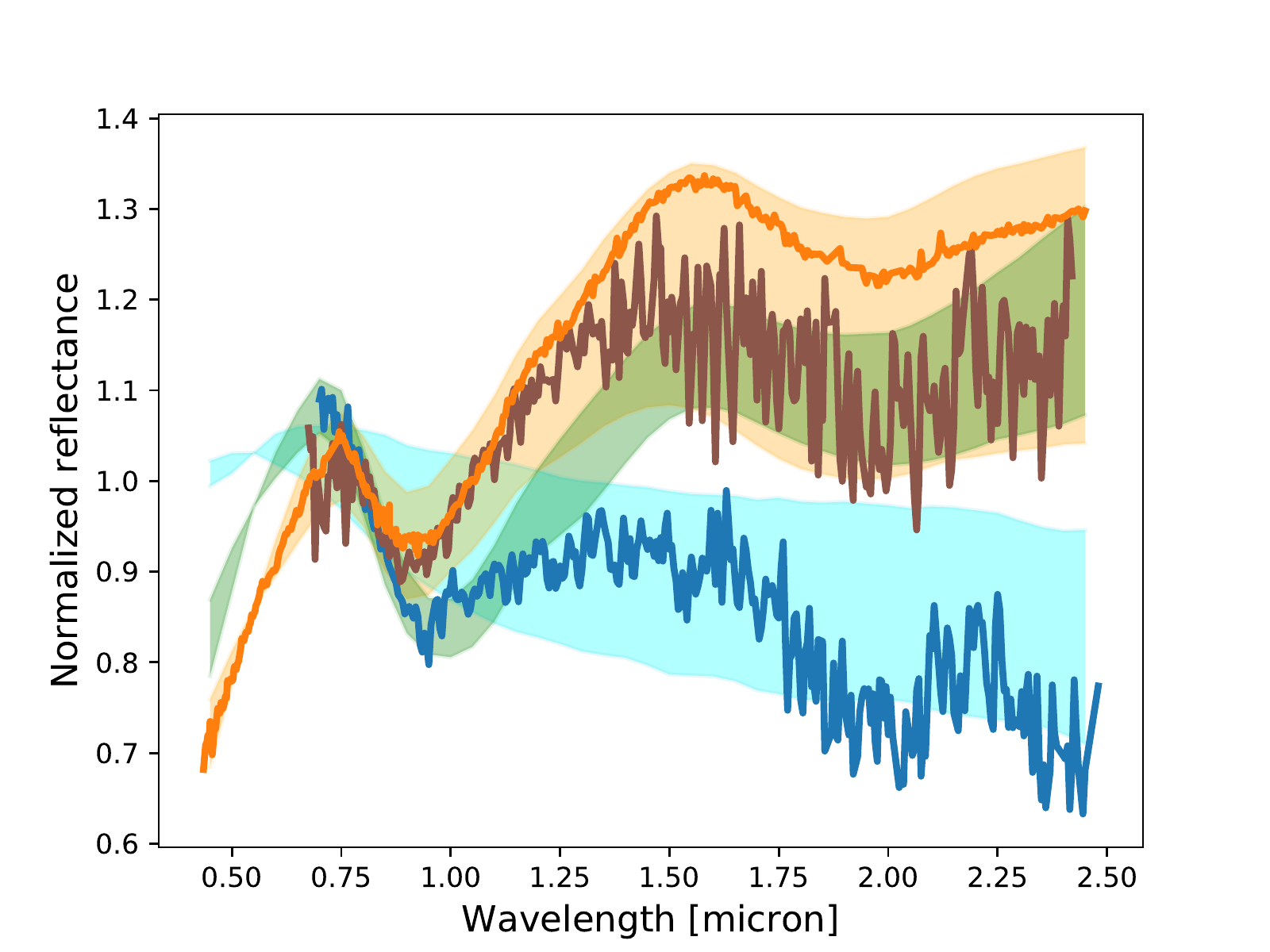}
   \caption{Spectral measurements of (6487)~Gault acquired on UT 2019 March 31 (blue) and UT 2019 April 8 (brown), compared to a previous measurement of (25)~Phocaea (orange; from the SMASS spectral database). All spectra are normalized to 1 at 0.8 $\upmu$m. 
   The orange, green, and blue regions indicate the mean ($\pm$1$\sigma$) spectral regions of S-type, Q-type, and B-type asteroids, respectively, from \citet{DeMeo:2009}.} 
\label{fig:spectra}
\end{figure}

\begin{deluxetable*}{lcccccc}
\tablecaption{Circumstances for spectroscopic and imaging observations of (6478)~Gault.}
\tablehead{
\colhead{UT Date} & \colhead{Mean MJD} & \colhead{Instrument} & \colhead{Air Mass} & \colhead{$r_H$ (au)} & \colhead{$\Delta$ (au)} & \colhead{$\alpha$ (deg)}}
\startdata
2019 Mar 31 & 58573.398553 & SpeX      & 1.09--1.33 & 2.30 & 1.44 & 16.0 \\
2019 Apr 8  & 58581.293721 & SpeX      & 1.05--1.06 & 2.28 & 1.48 & 18.9 \\
2019 Mar 23 & 58565.910038 & NEEMO T05 & 1.52--1.62 & 2.32 & 1.41 & 12.9 \\
2019 Mar 24 & 58566.849897 & NEEMO T05 & 1.45--1.47 & 2.31 & 1.41 & 13.3 \\
2019 Mar 30 & 58572.863234 & NEEMO T05 & 1.38--1.51 & 2.30 & 1.43 & 15.8 \\
2019 Mar 31 & 58573.849273 & NEEMO T05 & 1.38--1.46 & 2.30 & 1.44 & 16.1 \\
2019 Apr 02 & 58575.847994 & NEEMO T05 & 1.38--1.41 & 2.29 & 1.45 & 16.9 \\
2019 Apr 04 & 58577.821224 & NEEMO T05 & 1.36--1.37 & 2.29 & 1.46 & 17.7 \\
2019 Apr 24 & 58597.834428 & NEEMO T05 & 1.39--1.46 & 2.24 & 1.61 & 23.7 \\
2019 Apr 26 & 58599.782625 & NEEMO T05 & 1.27--1.30 & 2.24 & 1.63 & 24.1 \\
\enddata
 \tablecomments{We provide the date, mean Modified Julian Date (MJD), instrument, air mass, heliocentric ($r_H$) and topocentric ($\Delta$) distances, and phase angle ($\alpha$) of the asteroid.}
\label{tab:observations}
\end{deluxetable*}

Spectroscopic observations were conducted on two different nights with the 3-meter NASA Infrared Telescope Facility (IRTF) located on Maunakea, Hawaii. 
We used the SpeX NIR spectrograph \citep{Rayner:2003} combined with a 0.8$\times$15 arcsec slit in the low-resolution prism mode to measure the spectra over the 0.7--2.5~$\mu$m wavelength range. 
Series of spectral images with 120\,s exposure time were recorded in an AB beam pattern to allow efficient removal of the sky background by subtracting pairs of AB images. 
Three solar analog stars were observed throughout each night for data reduction purposes. 
We used \citet{Landolt:1973}'s stars 102-1081, 105-56 and 110-361 on the first night (UT 2019 March 31), and 98-978, 102-1081 and 105-56 on the second one (UT 2019 April 8). 
Details about the observing conditions are provided in Table~\ref{tab:observations}. 

Data reduction and spectral extraction followed the procedure outlined in \citet{Binzel2019}. 
We summarize it briefly here. 
Reduction of the spectral images was performed with the Image Reduction and Analysis Facility (IRAF) and Interactive Data Language (IDL), using the “autospex” software tool to automatically write sets of command files. 
Reduction steps for the science target and the calibration stars included trimming the images, creating a bad pixel map, flat-fielding the images, sky subtracting between AB image pairs, tracing the spectra in both the wavelength and spatial dimensions, co-adding the spectral images, extracting the 2-D spectra, performing wavelength calibration, and correcting for air mass differences between the asteroid and the solar analogs. 
Finally, the resulting asteroid spectrum was divided by the mean spectrum of the three solar analogs to remove the solar gradient.

\subsection{Spectrally variable Q-/S-type asteroid} \label{sec:results}

The SpeX NIR spectral measurements of (6478)~Gault are shown and compared to that of (25)~Phocaea 
in Fig.~\ref{fig:spectra}. 
Spectra from both nights of observation exhibit deep absorption bands near 1 and 2~$\mu$m consistent with an S- or Q-type surface composition \citep{Bus:2002, DeMeo:2009}. 
This result supports a compositional link between Gault and the Phocaea collisional family, and rules out an origin of Gault from the Tamara family, in agreement with results from other teams (\citealt{Sanchez:2019}, Licandro et al., private communication). 

Intriguingly, the measured spectral slope in our dataset drastically changed between the two nights of observations. 
On UT 2019 April 8, the spectrum was red over the 0.75--2.45\,$\upmu$m spectral range, with spectral gradient $s'$=9.1$\pm$1.2\% $\upmu$m$^{-1}$ (one standard deviation), which is consistent with most S-type asteroids, but lower than Phocaea ($s'$=22.0$\pm$0.2\% $\upmu$m$^{-1}$).  
On UT 2019 March 31, the spectrum was surprisingly blue, with negative (bluer-than-solar) spectral slope $s'$=--13.5$\pm$1.1\%\,$\upmu$m$^{-1}$. 
The only asteroid class defined by negative spectral slope is the B-type, but this class usually shows very little or no absorption features in the NIR. 
Here, the silicate absorption bands clearly indicate that the spectrum is compositionally consistent with the S-complex, although it is bluer than the vast majority of S-type and Q-type asteroids \citep{DeMeo:2009}. 

Several sanity checks were performed to confirm the blue spectral slope did not originate from any observational, instrumental or data reduction effect. 
Observations were collected in good weather and seeing ($0".4-0".8$) conditions, at low air mass (mostly $<$1.3) and low phase angle ($\alpha\,<\,20$\,${}^{\circ}$; Table~\ref{tab:observations}). 
We searched for possible flux calibration issues by producing a separate reflectance spectrum for each of the three solar analog stars observed during the night. 
We find that the spectral slope of the asteroid remains blue no matter which star is used, with little spectral variability of $\Delta s'$=2.7\%\,$\upmu$m$^{-1}$ across the three stars. 
Next, we compared other asteroids measured the same night with previous SpeX observations performed by our team. 
Already observed on UT 2019 January 25, asteroid (381677)~2009 BJ81 was targeted a few hours after (6487)~Gault on UT 2019 March 31 and showed no evidence for instrumental bluing during the night (the collected spectra can be retrieved from the SMASS database). 
It follows that if any instrumental or data reduction problem occurred, then it must have \textit{only} affected the exposures of Gault, which seems highly unlikely. 
To rule out this hypothesis, we divided our full sequence of spectral images at the sequence midpoint, extracted a 1D spectrum from each half sequence, and confirmed that Gault was blue during the two half-observing sequences. 

\section{Photometry} \label{sec:images}

\begin{figure*}
   \centering
   \includegraphics[width=0.85\linewidth, trim=0cm 0cm 0cm 0cm]{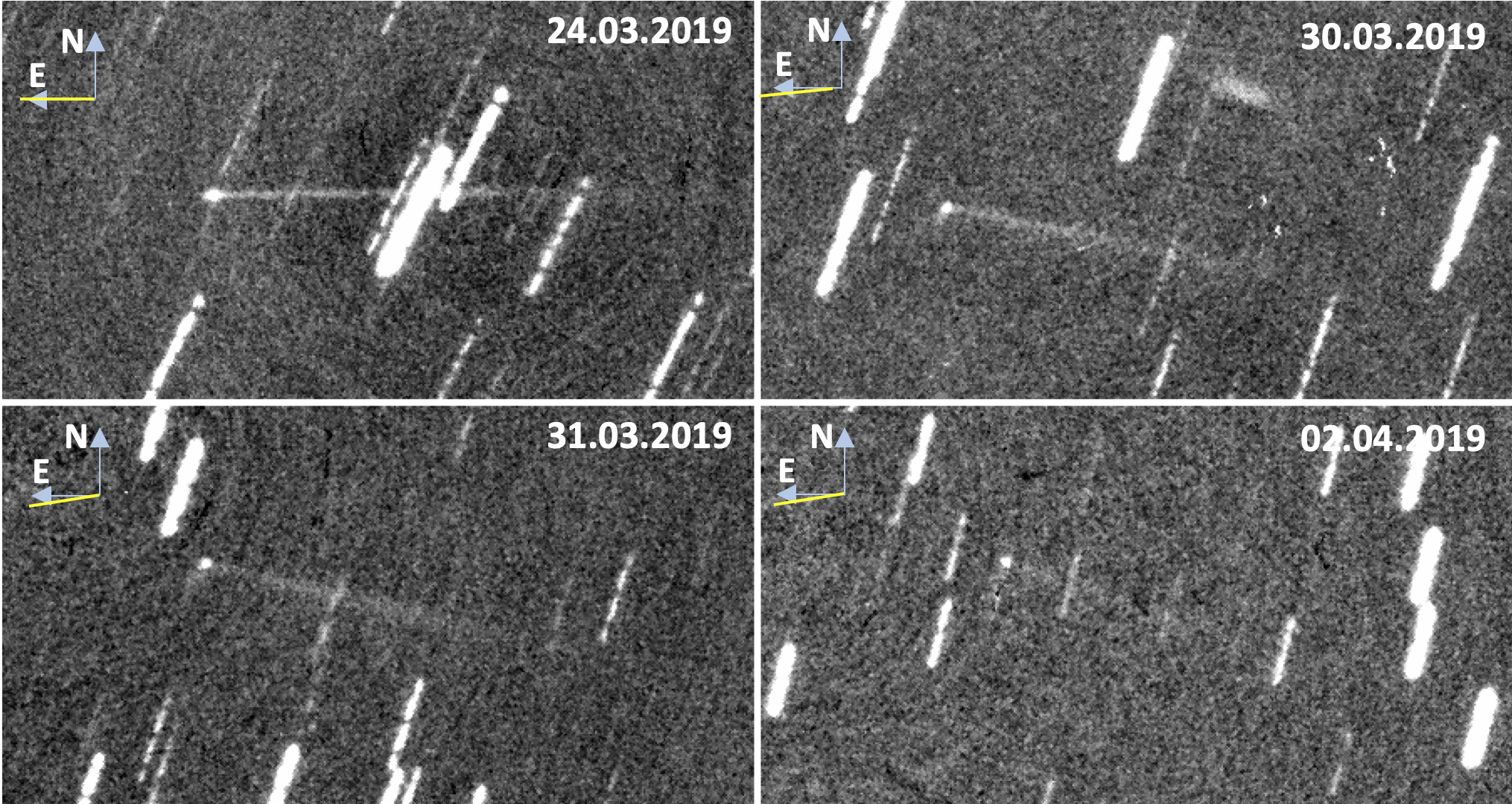} 
   \caption{Composite images of (6478)~Gault collected with the NEEMO~T05 telescope on four different nights of observation. In each image, the orientation of the field and the antisolar direction are indicated. Images were normalized to the same exposure time to show the apparent evolution of the tail as the Earth moved away from the orbital plane of the asteroid (plane-crossing date: UT 2019 March 24). \vspace{5mm}} 
\label{fig:imaging}
\end{figure*}

\subsection{Observations and data reduction} 

Nearly simultaneously to our spectroscopic observations, images of the asteroid were collected using the 0.5 m Near Earth Environment Monitoring (NEEMO) T05 telescope (RiDK500 f/7 manufacture) equipped with a FLI ProLine16803 4k$\times$4k camera \citep{2018RoAJ...28...67B, 2019RoAJ...B}. 
NEEMO~T05 is a relatively  compact telescope with a field of view of about 47$\times$47~arcmin installed inside the Astronomical Institute of the Romanian Academy (IAU code 073). 
120\,s long exposures with 2$\times$2 binning were measured in the non-filtered mode, meaning that the full 0.36$-$0.93\,${\rm \upmu m}$ wavelength range of the light was being recorded. 
Observing circumstances are summarized in Table~\ref{tab:observations}. 

The images were calibrated using flat and dark calibration files. 
Composite images built from shifted images centred on the asteroid are shown in Figure~\ref{fig:imaging}.
Differential aperture photometry was performed on the individual images with the MaximDL software,\footnote{http://diffractionlimited.com/product/maxim-dl/} using up to five reference stars with a signal-to-noise ratio that is greater than 100 for photometric calibration. 
We used an aperture of 4" and a sky annulus of 3", with a 2" gap between the two. 
Magnitudes of our reference stars were retrieved from the eight-filter (Johnson {\it B} and {\it V} and Sloan {\it u}, {\it g}, {\it r}, {\it i}, {\it z} and {\it Z}) AAVSO Photometric All-Sky Survey (APASS; \citealt{Henden:2015}). 
We used Johnson {\it V}-band magnitudes for calibration, which covers the closest spectral range to the non-filtered mode of the instrument. 
However, we acknowledge that the spectral response of our CCD sensor does not perfectly match Johnson {\it V}, thereby introducing a systematic error in our measurements. 
Finally, individual measurements taken on the same night were averaged together. 

\subsection{Photometric light curve}
\label{sec:lc}

Additional photometric measurements collected by other teams between 2019 January and May were used to complement our dataset. 
We used published data from the Xingming Observatory \citep{Hui:2019} graciously provided by the authors upon request, as well as measurements from the moving object detection system of the ZTF survey made available by \citet{Ye:2019b} on the Minor Planet Center. 

All photometric measurements were converted to absolute magnitudes to account for the changing geometry of the asteroid.
Observed magnitudes were reduced to heliocentric and geocentric distances $r_H\,=\,1\,{\rm au}$ and $\Delta\,=\,1\,{\rm au}$, and then corrected for phase angle using equation (2) in \citet{Hui:2019} for the compound (asteroid+tail) phase function.
We used \citet{bowell:1989}'s phase function for the asteroid, and \citet{Schleicher:1998}'s for dust in the tail. 
\citet{Ye:2019b} derived $H_{n,g}\,=\,14.81\,\pm\,0.04$ and $H_{n,r}\,=\,14.31\,\pm\,0.01$ for the asteroid based on pre-outburst photometric measurements of Gault from the ZTF survey assuming $G=0.15$. 
Here, we recalculated $H_{n,g}$ and $H_{n,r}$ from the same data for $G=0.23$, which is typical for S-complex asteroids (e.g., \citealt{Pravec:2012}). 
We find $H_{n,g}\,=\,14.88\,\pm\,0.04$ and $H_{n,r}\,=\,14.38\,\pm\,0.01$ which, converted to Johnson magnitudes using equations from \citet{Tonry:2012}, 
translates to $H_{n,B}=15.39\,\pm\,0.06$, $H_{n,V}=14.64\,\pm\,0.04$, $H_{n,R}=14.17\,\pm\,0.04$ . 
Finally, all absolute magnitudes were converted to a common Johnson {\it V}-band filter using average colors in the photometric dataset. 

\begin{figure*}
   \centering
   \includegraphics[width=\linewidth]{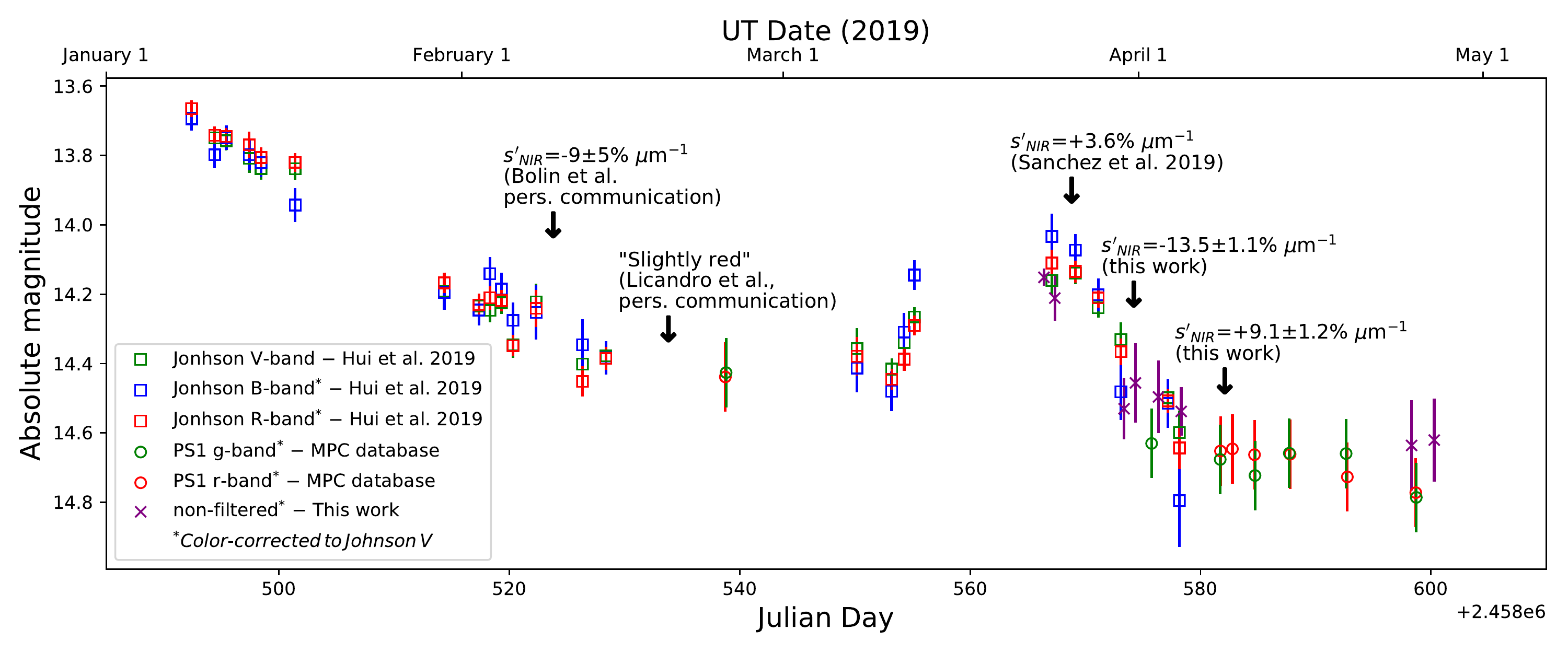}
   \caption{Absolute optical magnitude of (6478)~Gault between 2019 January and May, derived from photometric measurements collected by various research groups (see Section\,\ref{sec:lc}). All measurements were converted to Johnson {\it V}-band. The brightness peak near UT 2019 March 24 (JD=2,458,566.5) coincides with the asteroid plane-crossing time of Earth. Considering the fact that no new tail subsequently appeared, this brightening event was likely caused by the increased presence of dust from previous phases of activity in the orbital plane of Gault. NIR spectroscopic observations from our group and other teams are indicated on the figure (${s'}_{NIR}$ denotes the reported NIR spectral slope). Spectral slope does not appear to correlate with absolute magnitude. \vspace{1cm}
} 
\label{fig:photometry}
\end{figure*}

The resulting light curve is shown in Fig.~\ref{fig:photometry}: 
The asteroid brightness decreased from its phases of activity in late 2018/early 2019 up to the first half of March 2019.  
Transient brightening then occurred in March, peaking UT March 24 (JD=2,458,566.5), which coincides with the moment when the Earth crossed the orbital plane of the asteroid (one of asteroid orbital nodes). 
This indicates that brightening likely originated from the increased presence of dust in the orbital plane of the asteroid rather than an outburst, which is supported by the fact that no new tail subsequently appeared. 
No additional magnitude change indicative of an outburst is detected afterward. 

\section{Possible origins for spectral variability} \label{sec:discussion}

To further investigate the origin of NIR spectral variability in Gault, we searched for additional spectroscopic measurements collected by other research groups. 
\citet{Sanchez:2019} and Licandro et al. (private communication) measured red spectral slopes that are slightly more neutral than our second spectrum from UT 2019 April 8. 
Bolin et al. (private communication), on the other hand, derived a blue spectral slope of $s'\,=\,-9\,\pm\,5\%\,{\rm \upmu}$m from broadband {\it JHK} photometry, that is consistent with our first spectroscopic observation on UT 2019 March 31. 
This measurement was acquired about 50 days earlier than ours, implying that Gault underwent at least two distinct events of spectral bluing. 

Taken as a whole, these datasets do not reveal any correlation of NIR spectral slope with absolute magnitude (Fig.~\ref{fig:photometry}): color therefore does not appear to relate to the level of activity. 
Spectral bluing in particular does not coincide with local brightness minima or maxima, making it difficult to relate to either the surface or the tail of the asteroid. 

Various active comets are known to exhibit similar large-amplitude spectral variability in the optical, from blue to red colors, on temporally short timescales (e.g., \citealt{Weiler:2003,Ivanova:2017,Lukyanyk:2019}). 
The origin of variability in these objects remains unclear. Many different processes have been invoked, from the sublimation of icy grains, to physical and compositional variations of the emitted dust particles. 
In active S-type asteroids, the hypothesis of ice sublimation can likely be rejected. 
In our current understanding of the solar system formation and evolution, these objects accreted below the snow line \citep{vernazza:2017}, seemingly ruling out the presence of volatiles in their interior and, therefore, in the emitted dust. 
This is reinforced in the case of Gault by the lack of detectable gas in its visible spectrum \citep{Jewitt:2019} and the morphology of the tails that argue against sublimation as the driven mechanism for activity \citep{Ye:2019b}. 
Alternative hypothesis are as follows.

\begin{figure}
   \centering
   \includegraphics[width=8.5cm]{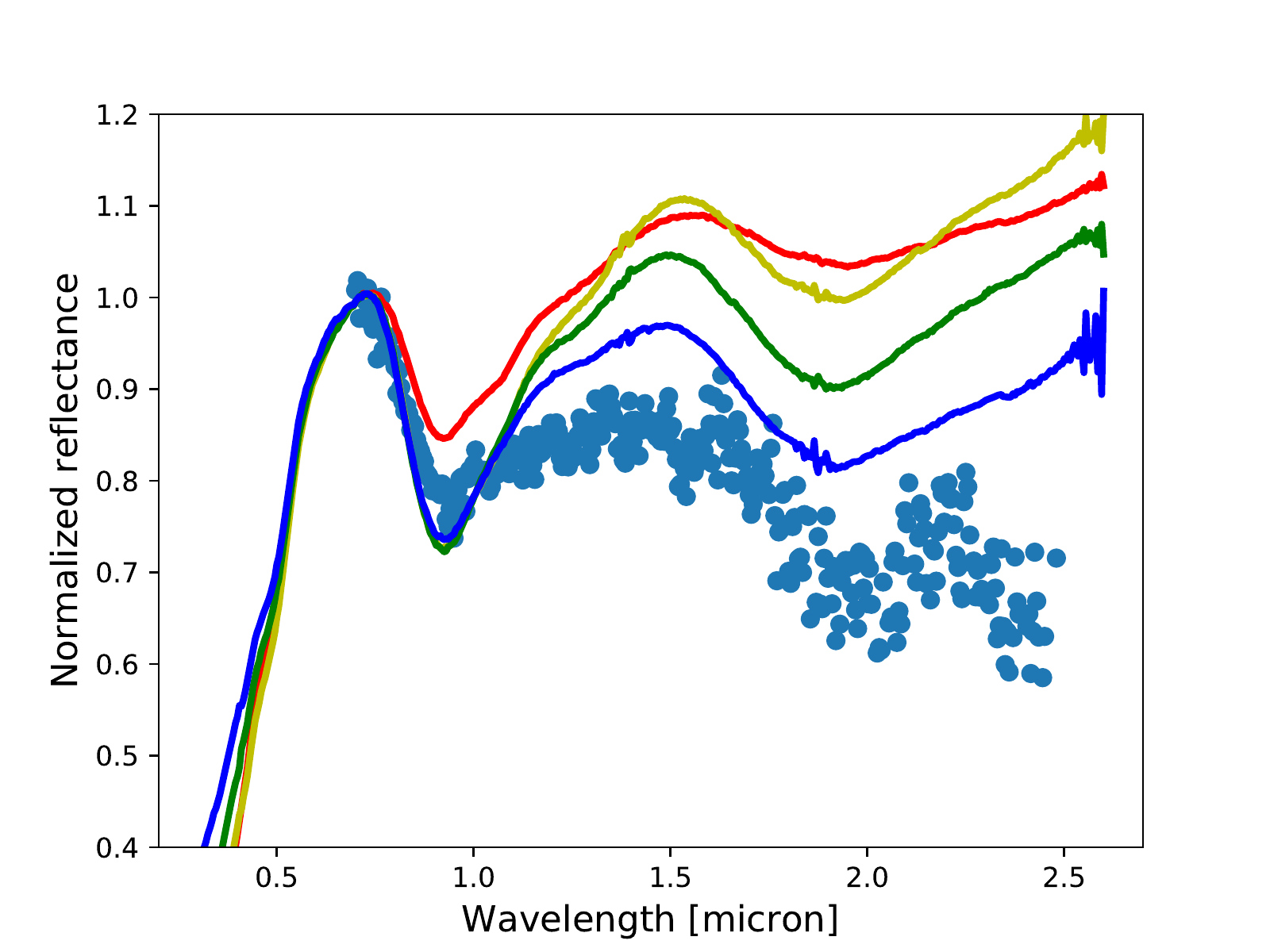}
   \includegraphics[width=8.5cm]{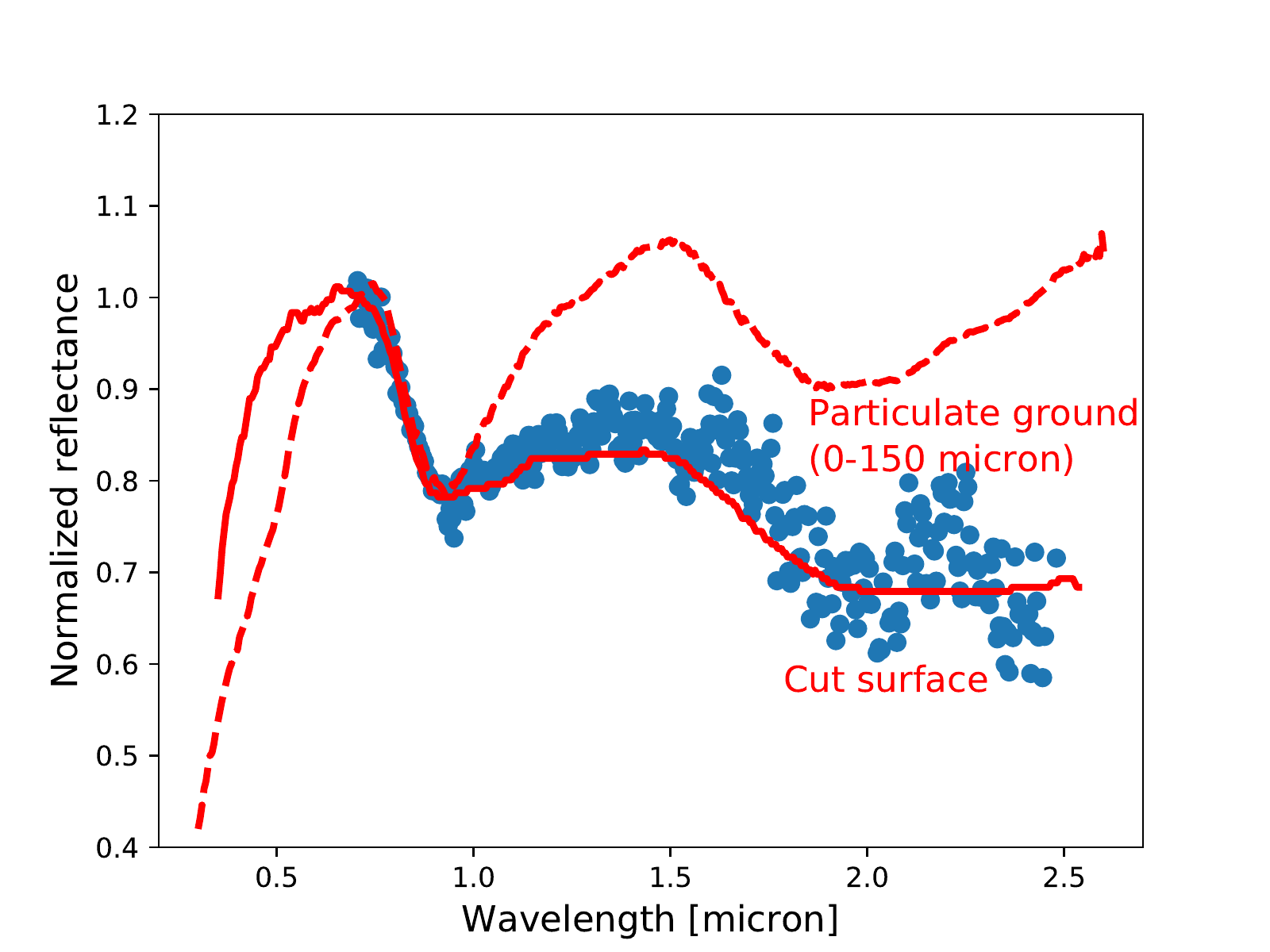}
\caption{Comparison between the blue spectrum of Gault collected on UT 2019 March 31 and laboratory samples of OC meteorites. 
Left panel: comparison with samples of the Yamato L3 OC meteorite with different grain sizes: 0--25\,$\upmu$m (red), 25--45\,$\upmu$m (yellow), 45--75\,$\upmu$m (green) and 75--125\,$\upmu$m (dark blue). OC grains samples show a trend of decreasing spectral slope with increasing grain size, but this effect is insufficient to account for the blue spectral slope of the asteroid. Right panel: comparison with a grained sample (red dashed line) and slab (continuous red line) of the Bald Mountain L4 OC. The meteorite slab provides the best spectral analog we could find in the RELAB database, opening the possibility that the asteroid blue color originated from partial loss of its dust regolith exposing regions of fresh, dust-free material at its surface.}
\label{fig:ocs}
\end{figure}

\begin{figure}
   \centering
   \includegraphics[width=6cm]{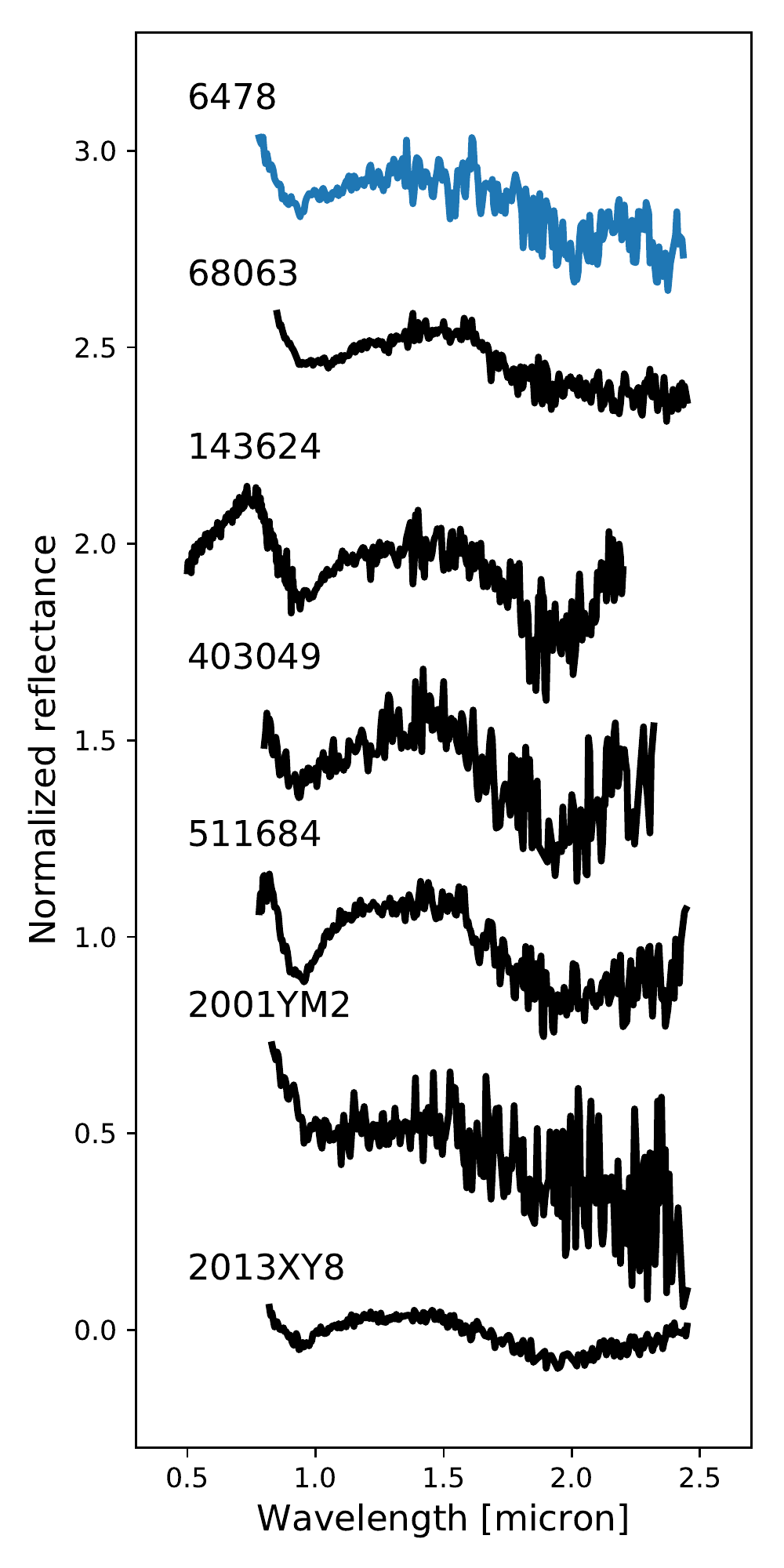}
   \caption{Spectral comparison between Gault (blue) and additional blue-slope Q-type asteroids from the SMASS database (black; see Table\,\ref{tab:blue_stypes}). 
   The presence of fast rotators among these objects supports the hypothesis of fresh dust-free surfaces as the origin of blue color among Q-type asteroids:
   (68063)~2000~YJ66 has a rotation period P=2.1\,h, which is close to the spin barrier. 
   2013~XY8 is a 30\,m-size extreme rotator with P=3.6\,minutes unable to sustain any dust regolith.}
\label{fig:stypes}
\end{figure}

\begin{enumerate}[A.]
 
     \item {\it Rayleigh scattering:} Non-geometrical scattering by optically thin particles has often been invoked as origin of the blue color of cometary comae. 
    However, while small particles certainly dominate by number in tails of active objects before being dispersed by solar wind, \citet{Jewitt:2015} argued that large grains should always dominate the scattering cross section. 
    In the case of Gault, this is further supported by the large average particle size (d$\sim$100\,$\upmu$m) derived from morphological analysis of the tails \citep{Jewitt:2019, Kleyna:2019, Moreno:2019, Ye:2019b}, implying that they do not contain a preponderance of Rayleigh scatterers. 
    
    \item {\it Size variation of the emitted dust grains:} S-complex bodies are associated to ordinary chondrite (OCs) meteorites \citep{Nakamura:2011}. Fresh (unweathered) samples of OCs 
    usually display rather neutral or slightly red NIR spectra. Like other meteorite types (e.g., \citealt{Cloutis:2011,Cloutis:2012}), OCs further exhibit a trend of decreasing spectral slope with increasing grain size: the coarser the grains, the more neutral the spectrum (Fig.~\ref{fig:ocs}). 
    We are not aware, however, of any spectrum of OC grains as blue as Gault on UT 2019 March 31. In particular, no matching spectrum of grained samples of OCs could be found in the RELAB spectral database of meteorites \citep{Pieters:2004}. 

    \item {\it Partially cleaned surface:} Dust-free slabs of OC meteorites provide the closest spectral analogs to the asteroid blue spectrum (Fig.\,\ref{fig:ocs}), opening the possibility that  
    a fresh blocky material exposed by the outbursts is responsible for the blue color. 
    Considering the fact that a red spectrum was measured at minimum brightness on UT 2019 April 8, when dust from the tail was only marginally contributing to the flux, the removal of the dust regolith must have been only partial on Gault, resulting in a spectrally heterogeneous surface. 
    If this hypothesis is correct, then additional Q-type bodies rotating close to break-up should exhibit the same blue spectral slope as Gault, as the result of recent centrifugal ejection of their dust regolith. 
    Indeed, some asteroid pairs, bodies that are suspected to be rotationally disintegrated asteroids \citep{Pravec:2010}, often present fresh spectral slopes \citep{Polishook:2014}. 
    A search for blue Q-type asteroids in the SMASS/MITHNEOS spectral database \citep{Binzel2019} reveals six bodies (Fig.~\ref{fig:stypes}), two of which are fast rotators near or beyond the spin limit (Table\,\ref{tab:blue_stypes}). The most extreme case is a 30-m size boulder with a 3.6\,minute long rotation period. 
    We acknowledge, however, that many other S-types that are not blue are also found near break-up speeds, implying that substantial regolith loss does not always happen on these objects. 
    The heterogeneous (partially cleaned) surface hypothesis therefore appears to us as the most credible explanation for spectral variability and transient bluing of Gault. 
    One possible flaw comes from the lack of spectral variability during our $\sim$50\,minute long sequences of spectroscopic observations. This is hardly reconcilable, in the case of an heterogeneous body, with the rotation period of $\sim$2\,hr proposed for Gault \citep{Kleyna:2019}. 

\end{enumerate}

\vspace{1cm}
\startlongtable
\begin{deluxetable}{llccc}
\tablecaption{Blue Q-type asteroids from the SMASS database
\label{tab:blue_stypes}}
\tablehead{
\colhead{Number} & \colhead{Designation} & \colhead{D$^a$ (km)} & \colhead{P$^{b,*}$ (h)} & Note}
\startdata
68063  & 2000~YJ66 & 2.36 & 2.11 & Binary \\
143624 & 2003 HM16 & 1.96 & 32.10 & -- \\
403049 & 2008~AY36  & \textit{unknown} & \textit{unknown} & -- \\
511684 & 2015~BN509  & 0.22 & 5.68 & -- \\
       & 2001~YM2  & \textit{unknown} & \textit{unknown} & -- \\
       & 2013~XY8  & 0.03 & 0.06 & -- \\
\enddata
\vspace{2mm}
{}$^a$Diameter, {}$^b$Rotation period, {}$^*$From \citealt{Warner:2009}
\end{deluxetable}

\section{Summary} \label{sec:summary}

We present new NIR spectroscopic observations and imaging of the newly active S-complex (Q- or S-type) asteroid (6478)~Gault. 
Combining our dataset with similar observations from other research groups, we showed that the asteroid exhibits substantial spectral variability, from very blue ($s'$=--13.5$\pm$1.1\%\,$\upmu$m$^{-1}$) to red ($s'$=+9.1$\pm$1.2\% $\upmu$m$^{-1}$) spectral slope over the 0.75--2.45\,$\upmu$m spectral range, that does not seem to correlate with activity. 
Spectral comparisons with samples of OC meteorites and other blue Q-type asteroids in the SMASS database suggest that the asteroid surface partially lost its dust regolith during the outburst events, thereby exposing a fresh blocky material at its surface. 
Future spectroscopic observations when the tails dissipate will help confirming or ruling out this hypothesis.

\section*{Acknowledgements}
We thank the anonymous referee for a careful reading of the manuscript and constructive remarks. 
This work is based on observations collected at the Infrared Telescope Facility, which is operated by the University of Hawaii under contract NNH14CK55B with the National Aeronautics and Space Administration. M.M. and F.D. were supported by the National Aeronautics and Space Administration under grant No. 80NSSC18K0849 issued through the Planetary Astronomy Program. 

\bibliographystyle{aasjournal.bst}
\bibliography{references}



\end{document}